\documentclass[prl,twocolumn,groupedaddress,showpacs]{revtex4}
\usepackage{color,graphics,amssymb,amsmath}

\newcommand{\fig}[1]{Fig.~\ref{#1}}
\newcommand{\eqn}[1]{Eq.~(\ref{#1})}

\newcommand{\ket}[1]{| #1 \rangle}

\newcommand{\intd}[1]{\int \!\! d #1 \ }
\newcommand{\be}{\begin{eqnarray}}
\newcommand{\ee}{\end{eqnarray}}

\newcommand{\ot}{\otimes}
\newcommand{\pauli}[4]{\left[\begin{array}{rr} #1&#2\\#3&#4 \end{array}\right]}

\begin{document}

\title{Phase estimation as a quantum nondemolition measurement}

\author{B.~C.~Travaglione}
 \email{btrav@physics.uq.edu.au}
\author{G.~J.~Milburn}
\author{T.~C.~Ralph}
 \affiliation{
  Centre for Quantum Computer Technology, University of Queensland,
  St. Lucia, Queensland, Australia
  }

\date{March 27, 2002}

\begin{abstract}

The phase estimation algorithm, which is at the heart of a variety of quantum
algorithms, including Shor's factoring algorithm, 
allows a quantum computer to accurately determine an eigenvalue of
an unitary operator. 
Quantum nondemolition measurements are a quantum mechanical procedure, used to
overcome the standard quantum limit when measuring an observable.
We show that the phase estimation algorithm, in both the discrete and
continuous variable setting, can be viewed as a quantum 
nondemolition measurement.

\end{abstract}

\maketitle


There is currently a great deal of active research aimed at developing
implementation schemes for quantum computation. For an overview of this field
see Nielsen and Chuang \cite{Nielsen00} or Preskill notes \cite{Preskill98}
and references therein.
The goal of quantum computation is to build a computer, which, using the laws of
quantum mechanics, can out perform any classical computer.
Using the laws of quantum mechanics to enhance a device's performance is not 
new.
In quantum optics, the procedure of quantum nondemolition (QND) measurement
\cite{Caves80,Milburn83,LaPorta89,Holland90,Poizat93}
aims to accurately measure an observable without perturbing it.
This procedure is consistent with the uncertainty principle, as, ideally, 
noise is induced only in the conjugate of the observable being measured.

Much of the impetus for research in quantum computation
stems from Shor's algorithm \cite{Shor94},
which allows a quantum computer to factor integers exponentially faster than 
can currently be done on a classical computer. 
The factoring of large integers allows the decryption of 
many public key encryption systems.
It has been shown that Shor's, and many other quantum
algorithms, can be described in terms of the phase estimation algorithm
\cite{Kitaev95,Cleve98}. 

It is evident that both phase estimation and QND measurement have the same
goal: they both aim to accurately measure the eigenvalue of an operator
(observable). In this paper we show that they are essentially the same
procedure. That is, every instance of the phase estimation algorithm can be
considered a QND measurement.

The original motivation for the study of QND measurements was the need to 
accurately measure weak classical forces \cite{Caves80}. 
Such measurements will be necessary, for example, in the detection of
gravitational radiation.
Classically, it is possible to measure both the position and momentum of a
particle with complete precision.
However, quantum mechanics tells us that the product of the uncertainty in the
position and momentum of a particle must be greater than Planck's constant
divided by $4\pi$,
\be\label{heisen1}
 \Delta x \Delta p &\geq& \frac{\hbar}{2}.
\ee
The consequence of the uncertainty principle for an oscillator is the fact
that any quantum mechanical state is characterized by an ``error box'' in the
phase plane with area at least
\be\label{heisen2}
 \Delta x \Delta p/m\omega &\geq& \hbar /2m\omega,
\ee
where $m$ and $\omega$ are the mass and angular frequency respectively
\cite{Caves80}. We can re-write \eqn{heisen2} in terms of the quadrature
variables, $\hat{X}$ and $\hat{Y}$,
\be\label{heisen3}
 \Delta X \Delta Y &\geq& \hbar /2m\omega,
\ee
where
\be
  \hat{x} + i\hat{p}/m\omega &=& (\hat{X} + i\hat{Y})e^{-i\omega t}.
\ee  
If we na\"{\i}vely perform an \emph{amplitude-and-phase} measurement of the 
quadrature variables, we obtain a round error box, with
\be
 \Delta X = \Delta Y &=& \sqrt{\frac{\hbar}{2m\omega}}.
\ee
However, by performing a back-action-evading QND measurement of, say,
$\hat{X}$, the error box becomes a long thin ellipse, with
\be
 \Delta X \ll \Delta Y
\ee
and $\Delta X$ can be made as small as one wishes, in principle.

Although a precise definition for a QND measurement is hard to pin down from
the literature, there are three basic criteria which a QND measurement must
satisfy. 
Firstly, it is required that the QND variable $\hat{A}$, which we wish to
measure, commutes with the free Hamiltonian evolution of the system,
\[
 \begin{array}{lcr}
 \hspace*{2.5cm} & [\hat{H}_{F}, \hat{A} ] = 0, & \hspace*{2cm} \mbox{(QND 1)}
 \end{array}
\]
so that if the system is measured in an eigenstate of $\hat{A}$, it remains in 
this eigenstate for all subsequent times, (perhaps only in the interaction 
picture).
The most important criterion which must be satisfied, is that the
Hamiltonian, $\hat{H}_{SM}$, coupling the meter variable $\hat{B}$ to 
the system must commute with the QND variable $\hat{A}$,
\[
 \begin{array}{lcr}
 \hspace*{2.5cm} & [\hat{H}_{SM}, \hat{A} ] = 0. & \hspace*{2cm} \mbox{(QND 2)}
 \end{array}
\]
This is known as the \emph{back action evasion criterion}.
Finally, it is also necessary that the meter variable $\hat{B}$, does
\emph{not} commute with the interaction Hamiltonian, 
\[
 \begin{array}{lcr}
 \hspace*{2.5cm} & [\hat{H}_{SM}, \hat{B} ] \ne 0. &\hspace*{2cm} \mbox{(QND 3)}
 \end{array}
\]
If this requirement is not fullfilled, then measuring the meter variable
$\hat{B}$ will yield no information about the QND variable $\hat{A}$.

To illustrate how a QND measurement procedure works, consider the example of
an ideal quadrature phase measurement of a single mode field \cite{Walls95}.
Suppose we have two degenerate modes $a$ and $b$, of an electromagnetic field.
We wish to measure $\hat{X}_a$ using the meter variable $\hat{X}_b$.
Now, if we assume that the amplitude
quadratures of each mode can be coupled according to the interaction 
Hamiltonian, 
\be\label{interact}
 \hat{H}_{SM} &=& \hbar \chi \hat{X}_a \hat{Y}_b, 
\ee
where $\chi$ is the coupling strength, then clearly (QND 2) and (QND 3) are 
satisfied. (QND 1) is also satisfied by choosing a rotating frame of
reference.
In order to draw analogies with the phase estimation algorithm, we will divide
the QND measurement into three steps. The first step involves initializing
mode $b$ into the zero eigenstate of the meter variable, $\ket{0}_b$,
\be
 \hat{X}_b \ket{0}_b &=& 0.
\ee
Now, for simplicity, we will assume that mode $a$ is already in an eigenstate 
$\ket{x}_a$ of the QND variable,
\be
 \hat{X}_a \ket{x}_a &=& x \ket{x}_a.
\ee 
Thus, the initial state of the quantum system is
\be
 \ket{x}_a \ket{0}_b.
\ee 
Re-writing the meter system in terms of the phase variable, $\hat{Y}_b$, gives
\be
 \intd{y} \ket{x}_a \ket{y}_b.
\ee
Evolving according to the interaction Hamiltonian (\eqn{interact}) results in
the state
\be
 e^{i\chi\hat{X}_a\hat{Y}_b} \intd{y} \ket{x}_a \ket{y}_b
  &=& \intd{y} e^{i\chi x y} \ket{x}_a \ket{y}_b.
\ee 
Finally re-writing mode $b$ in terms of the amplitude variable $\hat{X}_b$, we
have
\be
 \ket{x}_a \ket{\chi x}_b.
\ee
Therefore, by performing a quadrature measurement of the amplitude of mode $b$,
we obtain information about the state of mode $a$.

In the above description we have used eigenstates of the quadrature variables,
however, like position and momentum eigenstates, these are physically
unrealisable. However, they can be approximated to arbitrary degree by using
highly squeezed states.

The goal of the phase estimation algorithm is to obtain an eigenvalue 
of a unitary operator $\hat{U}$. 
Although hybrid systems have been discussed in the literature 
\cite{Travaglione01,Lloyd00}, generally the phase estimation algorithm 
is described in terms of (discrete) qubit registers \cite{Kitaev95,Cleve98}. 
Discrete classical computation, is accomplished through the manipulation of
binary digits (bits) using logic gates such as AND, OR, NOT and FANOUT,
whereas discrete quantum computation uses arrays of two level quantum systems 
(qubit registers) which are manipulated via one and two qubit unitary operators.
In analogy with the classical bit, it is conventional to use the symbols
$\ket{0}$ and $\ket{1}$ to denote the two levels of the quantum system.
Also in analogy with classical computation, we use the compact integer
representation to represent the basis states of a qubit register. For example,
\be\label{compeg}
 \ket{1}_3 \ot \ket{1}_2 \ot \ket{0}_1 \ot \ket{1}_0 &\equiv& \ket{13}.
\ee 
For an $m$ qubit register, we also introduce the Hermitian operator,
\be
 \hat{\Lambda} &=& \sum_{j=0}^{m-1} I \ot \cdots
 \ot I \ot \underbrace{2^{j-1}(I - Z)}_{j^\mathrm{th} \ \mathrm{term}}
 \ot I \ot \cdots \ot I,
\ee 
where $I$ is the identity operator, and $Z$ is the pauli operator, with matrix 
representation,
\be
 Z &=& \pauli{1}{0}{0}{-1}.
\ee 
The computational basis states are eigenstates of the operator $\hat{\Lambda}$ 
with an eigenvalue corresponding to their integer representation,
\be
 \hat{\Lambda}\ket{x} &=& x \ket{x} \quad \forall \ x \in [0,1,\dots,2^m-1].
\ee 
Another basis which is useful in understanding the phase estimation algorithm
is the Fourier transformed (FT) basis. This is the basis which arises by
performing the quantum Fourier tranform\cite{Coppersmith94}, $\hat{Q}$, on 
the computational basis,
\be
 \ket{\bar{x}} \equiv \hat{Q}\ket{x} \quad \forall \ x \in [0,1,\dots,2^m-1].
\ee
The Hermitian operator associated with this basis is
\be
 \hat{\Gamma} &=& \hat{Q}\hat{\Lambda}\hat{Q}^{-1},
\ee
as
\be
 \hat{\Gamma}\ket{\bar{x}} &=& x \ket{\bar{x}} 
             \quad \forall \ x \in [0,1,\dots,2^m-1].
\ee 

\begin{figure}[ht]
 \centering
 \scalebox{0.38}{\includegraphics{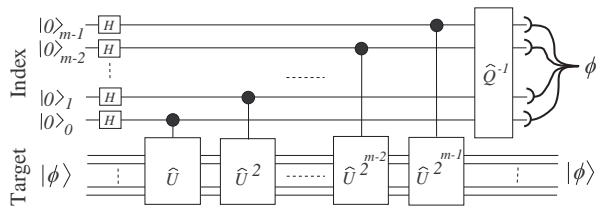}}
 \caption{Quantum circuit diagram representing the discrete phase estimation
 algorithm.}
 \label{peqnd1}
\end{figure}

Suppose that we have a register of qubits, the \emph{target} register, 
which is in an eigenstate $\ket{\phi}_T$, of some
unitary operator $\hat{U}$. We wish to determine the eigenvalue corresponding
to $\ket{\phi}_T$. As $\hat{U}$ is unitary, we have
\be\label{unitary}
  \hat{U}\ket{\phi}_T &=& e^{i\phi} \ket{\phi}_T ,
\ee
where $\phi$ is some number between 0 and $2\pi$.
The phase estimation algorithm utilizes an auxillary register of qubits, which
we shall call the \emph{index} register. 
\fig{peqnd1} depicts the quantum circuit diagram for the discrete 
phase estimation algorithm.
The algorithm begins by placing the index register in the zero of the FT basis,
\be
 \ket{\Phi_0} &=&  \ket{\bar{0}}_I\ket{\phi}_T,
\ee
Assuming that each qubit in the index register begins the computation in the
state $\ket{0}$, this step is accomplished by performing a Hadamard
gate, $\hat{H}$, on each qubit.
The next step is to perform a series of controlled $\hat{U}^{2^k}$ gates,
which couple the target and index registers.  
For each $k \in [0,1,\dots,m-1]$, the gate $\hat{U}^{2^k}$ is applied 
conditionally upon the state of the $k$-th qubit in the index register. 
The cummulative effect of applying these gates is the operation,
$\hat{U}^{\hat{\Lambda}}$.
The state of the system after the application of $\hat{U}^{\hat{\Lambda}}$ is
\be\label{phi1}
 \ket{\Phi_1} &=& \hat{U}^{\hat{\Lambda}}\ket{\Phi_0} \nonumber \\
          &=& \hat{U}^{\hat{\Lambda}}\ket{\bar{0}}_I\ket{\phi}_T \nonumber \\
 &=& \hat{U}^{\hat{\Lambda}} \frac{1}{\sqrt{M}} \sum_{j=0}^{M-1} \ket{j}_I \ket{\phi}_T \nonumber \\
 &=& \frac{1}{\sqrt{M}} \sum_{j=0}^{M-1} \ket{j}_I \hat{U}^j \ket{\phi}_T \nonumber \\
 &=&  \frac{1}{\sqrt{M}} \sum_{j=0}^{M-1} \ket{j}_I e^{i\phi j} \ket{\phi}_T,
\ee
where the last line makes use of \eqn{unitary}. The final step in the
algorithm involves measuring the index register in the FT basis, or
equivalently, performing the operation $\hat{Q}^{-1}$, and measuring in the
computational basis. This will, with high probability \cite{Travaglione01}, 
result in an $m$-bit approximation of $\phi$.

\begin{figure}[ht]
 \centering
 \scalebox{0.50}{\includegraphics{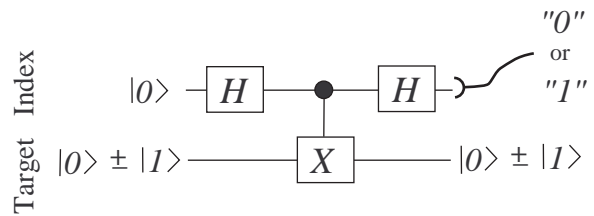}}
 \caption{A simple example of the phase estimation algorithm}
 \label{peqnd2}
\end{figure}

Having re-cast the phase estimation algorithm in terms of the FT basis, it is
now not difficult to see that the phase estimation algorithm is effectively a
QND measurement. We are trying to find an eigenvalue of the unitary operator,
$\hat{U}$, which can of course be written as 
\be
 \hat{U} &\equiv& e^{i\hat{H}_U},
\ee
for some Hermitian operator, $\hat{H}_U$.
This operator is the QND variable that we are trying to measure, $\hat{\Gamma}$
is the meter variable, and $\hat{H}_U\hat{\Lambda}$ is the interaction
Hamiltonian, as
\be\label{PEinteract}
 \hat{U}^{\hat{\Lambda}} &=& e^{i\hat{H}_U\hat{\Lambda}}.
\ee
In the quantum circuit model, it is assumed that the qubits undergo no free
evolution, so (QND 1) is automatically satisfied. It is clear from
\eqn{PEinteract} that the back action evasion criterion is satisfied
(QND 2), and it is not difficult to see that the interaction hamiltonian
does not commute with the meter variable (QND 3) as
\be
 [ \hat{\Lambda},\hat{\Gamma} ] &\ne& 0.
\ee
To illustrate how the phase estimation algorithm acts as a QND 
measurement, we describe the simplest possible algorithm. 
Suppose we wish to determine 
the eigenvalue of the pauli $X$ operator, or NOT gate,
\be
 X &=& \pauli{0}{1}{1}{0}.
\ee 
The eigenstates of this operator (ignoring a normalization factor) are 
$\ket{0} \pm \ket{1}$. The quantum circuit diagram for this simple phase 
estimation algorithm is depicted in \fig{peqnd2}.
In this example the QND variable is 
\be
 \hat{H}_U &=& \frac{\pi}{2}(I_T-X_T),
\ee 
the meter variable is 
\be
 \hat{\Gamma} = \hat{Q}\hat{\Lambda}\hat{Q}^{-1} 
        &=& \frac{H_I(I_I-Z_I)H_I}{2} = \frac{1}{2}\pauli{1}{-1}{-1}{1},
\ee
as the single qubit quantum Fourier transform is the self inverse Hadamard
operator, $H_I$, and the interaction Hamiltonian is
\be
 \hat{H}_U \hat{\Lambda} &=& \frac{\pi}{4}(I_T-X_T)(I_I-Z_I).  
\ee
Obviously, all the QND criteria are satisfied, and the measurement proceeds
by placing the meter system into the zero eigenstate of the meter variable,
$\ket{0} + \ket{1}$, which is accomplished by the first Hadamard gate, 
evolving according to the interaction Hamiltonian, which is accomplished by
the controlled NOT gate, and finally measuring the meter variable, by
performing another Hadamard and then measuring in the computational basis.

We have just shown that the discrete phase estimation algorithm is equivalent
to a QND measurement. 
Recently, a number of papers have been written exploring quantum computation 
using continuous variables \cite{Lloyd99,Gottesman01,Ralph01,Bartlett01}. 
We now show that continuous variable phase estimation is equivalent to
continuous variable QND measurement.

As with all quantum algorithms, the phase estimation algorithm can be
partitioned into three stages; the \emph{initialization} stage, the
\emph{entangling} stage and the \emph{measurement} stage.
The initialization stage involves placing the index system into the zero 
eigenstate of the FT basis. The entangling stage
involves the application of the operator
\be
 \hat{U}^{\hat{\Lambda}} &=& e^{i\hat{H}_U\hat{\Lambda}},
\ee 
and the measurement stage involves measuring the index system in the conjugate
of the computational basis.
If we replace the discrete qubit registers with infinite level (continuous)
quantum variables, and carry out each of these steps, we obtain 
something very similar to the ideal quadrature QND measurement discussed
earlier.
Suppose that the continuous variable operator, acting on the target system,
whose eigenvalue we wish to determine is $e^{i\hat{A}_T}$, 
and we choose the position of the continuous
index system as the computational basis. Then following the steps of the
phase estimation algorithm, we prepare our index system in the zero  
eigenstate of the conjugate basis, that is, the zero momentum eigenstate. 
This step corresponds to initialising our meter variable.
Then we perform the $\hat{U}$ operator, conditioned upon the computational 
basis, $e^{i\hat{x}_I\hat{A}_T}$, which corresponds to applying an
interaction Hamiltonian which satisfies (QND 1) and (QND 2). Before finally
performing a Fourier transform and measuring the position variable, which is 
equivalent to measuring the momentum of the index system (i.e. the meter 
variable). This is clearly a QND measurement of $\hat{A}$.

Of course, position eigenstates are physically unrealisable, 
and any continuous variable
implementation of the phase estimation algorithm would need to be done using
approximate position eigenstates, such as squeezed states. Just as ideal QND
measurements can only be approximated experimentally, ideal continuous
variable phase estimation can also only be approximated.

The study of quantum computation brings together a number of disciplines
including computer science, mathematics and quantum mechanics.
We have taken the phase estimation algorithm, which has its roots in computer
science, and shown that it is equivalent to the quantum mechanical procedure of
QND measurement. We have also described a continuous variable version of the 
phase estimation algorithm. 
It is interesting to speculate on whether other algorithms such as Grover's
algorithm might have similarities with well studied quantum phenomenon.

\acknowledgments

BCT and GJM thank S. Lloyd for helpful discussions. 


\begin{thebibliography}{18}
\expandafter\ifx\csname natexlab\endcsname\relax\def\natexlab#1{#1}\fi
\expandafter\ifx\csname bibnamefont\endcsname\relax
  \def\bibnamefont#1{#1}\fi
\expandafter\ifx\csname bibfnamefont\endcsname\relax
  \def\bibfnamefont#1{#1}\fi
\expandafter\ifx\csname citenamefont\endcsname\relax
  \def\citenamefont#1{#1}\fi
\expandafter\ifx\csname url\endcsname\relax
  \def\url#1{\texttt{#1}}\fi
\expandafter\ifx\csname urlprefix\endcsname\relax\def\urlprefix{URL }\fi
\providecommand{\bibinfo}[2]{#2}
\providecommand{\eprint}[2][]{\url{#2}}

\bibitem[{\citenamefont{Nielsen and Chuang}(2000)}]{Nielsen00}
\bibinfo{author}{\bibfnamefont{M.~A.} \bibnamefont{Nielsen}} \bibnamefont{and}
  \bibinfo{author}{\bibfnamefont{I.~L.} \bibnamefont{Chuang}},
  \emph{\bibinfo{title}{Quantum Computation and Quantum Information}}
  (\bibinfo{publisher}{Cambridge University Press},
  \bibinfo{address}{Cambridge}, \bibinfo{year}{2000}).

\bibitem[{\citenamefont{Preskill}(1998)}]{Preskill98}
\bibinfo{author}{\bibfnamefont{J.}~\bibnamefont{Preskill}},
  \emph{\bibinfo{title}{Quantum Information and Computation}},
  \bibinfo{organization}{California Institute of Technology},
  \bibinfo{address}{Pasadena, CA, USA} (\bibinfo{year}{1998}).

\bibitem[{\citenamefont{Caves et~al.}(1980)\citenamefont{Caves, Thorne, Drever,
  Sandberg, and Zimmermann}}]{Caves80}
\bibinfo{author}{\bibfnamefont{C.~M.} \bibnamefont{Caves}},
  \bibinfo{author}{\bibfnamefont{K.~S.} \bibnamefont{Thorne}},
  \bibinfo{author}{\bibfnamefont{R.~W.~P.} \bibnamefont{Drever}},
  \bibinfo{author}{\bibfnamefont{V.~D.} \bibnamefont{Sandberg}},
  \bibnamefont{and}
  \bibinfo{author}{\bibfnamefont{M.}~\bibnamefont{Zimmermann}},
  \bibinfo{journal}{Reviews of Modern Physics} \textbf{\bibinfo{volume}{52}},
  \bibinfo{pages}{341} (\bibinfo{year}{1980}).

\bibitem[{\citenamefont{Milburn and Walls}(1983)}]{Milburn83}
\bibinfo{author}{\bibfnamefont{G.~J.} \bibnamefont{Milburn}} \bibnamefont{and}
  \bibinfo{author}{\bibfnamefont{D.~F.} \bibnamefont{Walls}},
  \bibinfo{journal}{Physical Review A} \textbf{\bibinfo{volume}{28}},
  \bibinfo{pages}{2065} (\bibinfo{year}{1983}).

\bibitem[{\citenamefont{LaPorta et~al.}(1989)\citenamefont{LaPorta, Slusher, and
  Yurke}}]{LaPorta89}
\bibinfo{author}{\bibfnamefont{A.} \bibnamefont{LaPorta}},
  \bibinfo{author}{\bibfnamefont{R.~E.} \bibnamefont{Slusher}},
  \bibnamefont{and} \bibinfo{author}{\bibfnamefont{B.}~\bibnamefont{Yurke}},
  \bibinfo{journal}{Physical Review Letters} \textbf{\bibinfo{volume}{62}},
  \bibinfo{pages}{28} (\bibinfo{year}{1989}).

\bibitem[{\citenamefont{Holland et~al.}(1990)\citenamefont{Holland, Collett,
  Walls, and Levenson}}]{Holland90}
\bibinfo{author}{\bibfnamefont{M.~J.} \bibnamefont{Holland}},
  \bibinfo{author}{\bibfnamefont{M.~J.} \bibnamefont{Collett}},
  \bibinfo{author}{\bibfnamefont{D.~F.} \bibnamefont{Walls}}, \bibnamefont{and}
  \bibinfo{author}{\bibfnamefont{M.~D.} \bibnamefont{Levenson}},
  \bibinfo{journal}{Physical Review A} \textbf{\bibinfo{volume}{42}},
  \bibinfo{pages}{2995} (\bibinfo{year}{1990}).

\bibitem[{\citenamefont{Poizat and Grangier}(1993)}]{Poizat93}
\bibinfo{author}{\bibfnamefont{J.~P.} \bibnamefont{Poizat}} \bibnamefont{and}
  \bibinfo{author}{\bibfnamefont{P.}~\bibnamefont{Grangier}},
  \bibinfo{journal}{Physical Review Letters} \textbf{\bibinfo{volume}{70}},
  \bibinfo{pages}{271} (\bibinfo{year}{1993}).

\bibitem[{\citenamefont{Shor}(1994)}]{Shor94}
\bibinfo{author}{\bibfnamefont{P.~W.} \bibnamefont{Shor}},
  \bibinfo{journal}{Proc. 35th Annual Symposium on Foundations of Computer
  Science} p. \bibinfo{pages}{124} (\bibinfo{year}{1994}).

\bibitem[{\citenamefont{Kitaev}()}]{Kitaev95}
\bibinfo{author}{\bibfnamefont{A.~Y.} \bibnamefont{Kitaev}},
  \bibinfo{note}{quant-ph/9511026}.

\bibitem[{\citenamefont{Cleve et~al.}(1998)\citenamefont{Cleve, Ekert,
  Macchiavello, and Mosca}}]{Cleve98}
\bibinfo{author}{\bibfnamefont{R.}~\bibnamefont{Cleve}},
  \bibinfo{author}{\bibfnamefont{A.}~\bibnamefont{Ekert}},
  \bibinfo{author}{\bibfnamefont{C.}~\bibnamefont{Macchiavello}},
  \bibnamefont{and} \bibinfo{author}{\bibfnamefont{M.}~\bibnamefont{Mosca}},
  \bibinfo{journal}{Proc. Roy. Soc. London A} \textbf{\bibinfo{volume}{454}},
  \bibinfo{pages}{339} (\bibinfo{year}{1998}).

\bibitem[{\citenamefont{Walls and Milburn}(1995)}]{Walls95}
\bibinfo{author}{\bibfnamefont{D.~F.} \bibnamefont{Walls}} \bibnamefont{and}
  \bibinfo{author}{\bibfnamefont{G.~J.} \bibnamefont{Milburn}},
  \emph{\bibinfo{title}{Quantum Optics}} (\bibinfo{publisher}{Springer-Verlag},
  \bibinfo{address}{Berlin}, \bibinfo{year}{1995}).

\bibitem[{\citenamefont{Travaglione and Milburn}(2001)}]{Travaglione01}
\bibinfo{author}{\bibfnamefont{B.~C.} \bibnamefont{Travaglione}}
  \bibnamefont{and} \bibinfo{author}{\bibfnamefont{G.~J.}
  \bibnamefont{Milburn}}, \bibinfo{journal}{Physical Review A}
  \textbf{\bibinfo{volume}{63}}, \bibinfo{pages}{032301}
  (\bibinfo{year}{2001}).

\bibitem[{\citenamefont{Lloyd}(2000)}]{Lloyd00}
\bibinfo{author}{\bibfnamefont{S.}~\bibnamefont{Lloyd}},
  \emph{\bibinfo{title}{Hybrid quantum computing}} (\bibinfo{year}{2000}),
  \bibinfo{note}{to be published}.

\bibitem[{\citenamefont{Coppersmith}(1994)}]{Coppersmith94}
\bibinfo{author}{\bibfnamefont{D.}~\bibnamefont{Coppersmith}}
  (\bibinfo{year}{1994}), \bibinfo{note}{{I}BM Research Report No. RC19642}.

\bibitem[{\citenamefont{Lloyd and Braunstein}(1999)}]{Lloyd99}
\bibinfo{author}{\bibfnamefont{S.}~\bibnamefont{Lloyd}} \bibnamefont{and}
  \bibinfo{author}{\bibfnamefont{S.~L.} \bibnamefont{Braunstein}},
  \bibinfo{journal}{Physical Review Letters} \textbf{\bibinfo{volume}{82}},
  \bibinfo{pages}{1784} (\bibinfo{year}{1999}).

\bibitem[{\citenamefont{Gottesman et~al.}(2001)\citenamefont{Gottesman, Kitaev,
  and Preskill}}]{Gottesman01}
\bibinfo{author}{\bibfnamefont{D.}~\bibnamefont{Gottesman}},
  \bibinfo{author}{\bibfnamefont{A.}~\bibnamefont{Kitaev}}, \bibnamefont{and}
  \bibinfo{author}{\bibfnamefont{J.}~\bibnamefont{Preskill}},
  \bibinfo{journal}{Physical Review A} \textbf{\bibinfo{volume}{64}},
  \bibinfo{pages}{012310} (\bibinfo{year}{2001}).

\bibitem[{\citenamefont{Ralph et~al.}(2001)\citenamefont{Ralph, Munro, and
  Milburn}}]{Ralph01}
\bibinfo{author}{\bibfnamefont{T.~C.} \bibnamefont{Ralph}},
  \bibinfo{author}{\bibfnamefont{W.~J.} \bibnamefont{Munro}}, \bibnamefont{and}
  \bibinfo{author}{\bibfnamefont{G.~J.} \bibnamefont{Milburn}},
  \emph{\bibinfo{title}{Quantum computation with coherent states, linear
  interactions and superposed resources}} (\bibinfo{year}{2001}),
  \bibinfo{note}{quant-ph/0110115}.

\bibitem[{\citenamefont{Bartlett and Sanders}(2001)}]{Bartlett01}
\bibinfo{author}{\bibfnamefont{S.~D.} \bibnamefont{Bartlett}} \bibnamefont{and}
  \bibinfo{author}{\bibfnamefont{B.~C.} \bibnamefont{Sanders}},
  \emph{\bibinfo{title}{Universal continuous-variable quantum computation:
  Requirement of optical nonlinearity for photon counting}}
  (\bibinfo{year}{2001}), \bibinfo{note}{quant-ph/0110039}.

\end{thebibliography}




\end{document}